\documentclass[onecolumn,secnumarabic,amssymb,
amsmath,nobibnotes, aps, showpacs,prl]{revtex4}

\def\be{\begin{equation}}
\def\ee{\end{equation}}
\def\ba{\begin{eqnarray}}
\def\ea{\end{eqnarray}}

\newcommand{\beqa}{\begin{eqnarray}}
\newcommand{\eeqa}{\end{eqnarray}}

\newcommand{\bn}{\hat{\bf n}}
\newcommand{\bl}{\hat{\bf l}}
\newcommand{\beq}{\begin{equation}}
\newcommand{\eeq}{\end{equation}}
\newcommand{\bfl}{{\mathbf{l}}}

\newcommand{\bfL}{{\mathbf{L}}}
\newcommand{\bfLp}{{\mathbf{L^{\prime}}}}
\newcommand{\bflp}{{\mathbf{l^{\prime}}}}

\newcommand{\intl}[1]{\int {d^2 l_{#1} \over (2\pi)^2}}
\newcommand{\intlp}[1]{\int {d^2 l_{#1}' \over (2\pi)^2}}

\newcommand{\calb}{a}
\newcommand{\rot}{\omega}

\newcommand{\cmb}{T}

\newlength{\tskip}
\newlength{\colwidth}

\usepackage{color}
\usepackage{graphicx}
\usepackage{bm}

\def\ea{\ et al. \,}

\def\be{\begin{equation}}
\def\ee{\end{equation}}
\def\bea{\begin{eqnarray}}
\def\eea{\end{eqnarray}}

\newcommand{\apjs}{ApJS}

\begin{document}
                                                                                
\title{Revealing Cosmic Rotation}

\author{Amit P.S. Yadav$^{1,3}$, Meir Shimon$^{2}$, and Brian G. Keating$^3$}
\affiliation{$^1$Institute for Advanced Study, School of Natural Sciences, Einstein Drive, Princeton, NJ 08540}
\affiliation{$^2$School of Physics and Astronomy, Tel Aviv University, Tel Aviv 69978, Israel}
\affiliation{$^3$Center for Astrophysics and Space Sciences, University of California, San Diego, 9500 Gilman Drive, La Jolla, CA, 92093-0424}

\begin{abstract}
Cosmological Birefringence (CB), a rotation of the polarization plane of radiation coming to us from distant astrophysical sources, may reveal parity violation in either the electromagnetic 
or gravitational sectors of the fundamental interactions in nature. Until only 
recently this phenomenon could be probed with only radio observations or observations 
at UV wavelengths. Recently, there is a substantial effort to constrain such non-standard models 
using observations of the rotation of  the polarization plane of cosmic 
microwave background (CMB) radiation. This can be done via measurements of the $B$-modes of the CMB or by measuring its  
$TB$ and EB correlations which vanish in the standard model. In this paper we show that $EB$ correlations-based estimator is the best for upcoming polarization experiments. The $EB$ based estimator surpasses other estimators because it has the smallest noise and of all the estimators is least affected by systematics.   Current polarimeters are optimized for the detection of $B$-mode polarization 
from either primordial gravitational waves or by large scale structure via gravitational lensing. In the paper we also study optimization of 
CMB experiments for the detection of cosmological birefringence, in the presence of 
instrumental systematics, which by themselves are capable of producing $EB$ 
correlations; potentially mimicking CB. 
\end{abstract}


\date{\today}

\maketitle
                                                                



\maketitle
        \section{Introduction}

The cosmic microwave background (CMB) is, arguably, the ideal probe of the standard cosmological 
model. The polarization of the CMB can be studied in terms of the parity-even $E$ and parity-odd 
$B$-modes~\cite{SZ97,SZ98,1997PhRvD..55.7368K,1997PhRvL..78.2058K}.
In the standard cosmological model, the physics governing the radiating field is parity 
invariant. Hence, the parity odd correlations
$\langle T B \rangle$, $\langle E B \rangle$ vanish identically. However, the plane of the CMB's linear polarization can be 
rotated due to interactions which introduce different 
dispersion relations for  left and right circularly 
polarized modes, during propagation to us from the last 
scattering surface. Such rotations generate non-zero $\langle T B \rangle$ and $\langle E B \rangle$cross-correlations in the CMB.
Thus, measurement of these correlations allow estimation of the 
rotation of the plane of the CMB polarization~\citep{1999PhRvL..83.1506L}. Such rotation can come from several processes/sources: \emph{e.g.,} foregrounds, Faraday rotation 
due to interactions with magnetic fields, and interactions with 
pseudoscalar fields~\cite{Carroll98}. The interaction with foregrounds and Faraday rotation lead to frequency dependent effects; the latter having a frequency dependence ($\propto \nu^{-2}$) ~\citep{Kosowsky_Loeb1996,2005PhRvD..71d3006K, 2004ApJ...616....1C,2004PhRvD..70f3003S}, while  interactions with pseudo-scalar fields are usually assumed to be frequency
independent. The distinct frequency dependencies allow separation of these effects.


We know that parity is violated by weak interactions and is possibly violated in the early universe, giving rise to baryon asymmetry. Hence, investigating the existence of parity violating interactions involving 
cosmologically evolving scalar fields is well-motivated.  As an example, an interaction of the form $\frac{\phi}{2M}F_{\mu \nu}{\tilde F}^{\mu \nu}$~\cite{Carroll98,prs08},  rotates the polarization plane of linearly polarized light by an angle of rotation 
$\alpha = \frac{1}{M}\int d\tau \dot{\phi}$ 
during propagation for a conformal time $\tau$. Here $F_{\mu\nu}$ is electromagnetic  strength tensor, and $\tilde F^{\mu\nu}$ is its dual.  The fluctuations in the scalar field $\phi$ are then encoded in the rotation angle $\alpha$ of the polarization.

Faraday rotation (FR), an interaction of CMB with magnetic fields, rotates the plane of polarization by angle $\alpha = \frac{3}{{16 \pi^2 e}} \lambda_0^2 
\int \dot{\tau} \ {\bf B} \cdot d {\bf l} \ ,$
where $\dot{\tau} \equiv n_e \sigma_T a$ is the differential 
optical depth, $n_e$ is the line of sight free electron density, 
$\sigma_T$ is the Thomson scattering cross-section, $a$ is the scale factor, 
$\lambda_0$ is the observed wavelength of the radiation, ${\bf B}$ is
the ``comoving'' magnetic field, and $d{\bf l}$ is the comoving
length element along the photon trajectory. Magnetic fields are prevalent in cosmic structures at high redshift~\cite{2008Natur.455..638W} and it is possible that they may have generated from primordial seed fields imprinted in the early universe (see \cite{Grasso_Rubinstein_2001} for a review).  It has been shown that constraining FR using the CMB polarization information is a leading diagnostic of primordial magnetic field~\cite{2012arXiv1207.3356Y, 2011arXiv1106.1438P}. 

As we will show, upcoming CMB polarization probes have the potential constrain the CB rotation angle, $\alpha$, to unprecedented precision -- at the $1'$ level. The objective of this paper is to seek optimization schemes for a family of proposed ground-based CMB experiments to detect cosmological birefringence.  In particular, we consider the possibility of increasing the size of observed sky patch 
at the expense of increasing the map noise of the experiments and explore how 
this may affect the bounds on cosmic birefringence (CB) that these experiments set.  We have considered a range of experiments (varying from Planck-like to cosmic-variance-limited experiment up to $\ell=3000$) to study general trends.

\section{The CMB and Cosmic Birefringence}

In the standard model $E$ and $B$ are pure parity states (even and odd, respectively). 
The correlation, over the full sky, of the $B$-mode with either the temperature or 
$E$-mode polarization vanishes in this case. However in the presence of CB the polarization plane is 
rotated, generating $E-B$ mixing which induces ``forbidden" $TB$ and $EB$ power spectra. The un-rotated CMB temperature field and the Stokes parameters  at angular position $\bn$  are written as
$\tilde{T}(\bn)$, and $\tilde{Q}(\bn)$, $\tilde{U}(\bn)$, respectively.
The temperature field is invariant 
under a rotation of the polarization by an angle $\alpha(\bn)$ 
at the angular position $\bn$,
while the Stokes parameters transform like a spin-2 field:
\begin{equation}
(Q(\bn) \pm iU(\bn))
=(\tilde{Q}(\bn) \pm \tilde{U}(\bn)) \exp(\pm 2i\alpha(\bn))
\label{RotationTransformationOfStokes}
\end{equation}
 
The $E$ and $B$ fields of the CMB are constructed from 
 observed Stokes parameters. In a Fourier basis (in the flat-sky approximation),
\begin{eqnarray}
 \left[ E\pm i B \right] (\bfl) &=&
        \int  d \bn\, [Q(\bn)\pm i U(\bn)] e^{\mp 2i\varphi_{\bf l}} e^{-i \bl \cdot \bn}\,,
\label{EBFields}
\end{eqnarray}
where $\varphi_{\bfl}=\cos^{-1}(\hat {\bf n} \cdot \hat \bfl)$. 
The change in the CMB fields due to rotation is
\begin{eqnarray}
\delta T(\bfl) &=& 0 \,,\\
\delta B(\bfl)      &=& 2\intlp{}
\Big[ \tilde E(\bfl') \cos 2\varphi_{\bfl'\bfl}
     -\tilde B(\bfl') \sin 2\varphi_{\bfl'\bfl} \Big] \alpha(\bfL)
,\nonumber\\
\delta E(\bfl)      &=& -2\intlp{}
\Big[ \tilde B(\bfl') \cos 2\varphi_{\bfl'\bfl}
     +\tilde E(\bfl') \sin 2\varphi_{\bfl'\bfl}\Big] \alpha(\bfL),\nonumber
\end{eqnarray}
where $\bfL=\bfl - \bflp$, and
$\varphi_{\bfl \bflp}= \varphi_\bfl - \varphi_\bflp$.
Thus, due to rotation, a mode of wavevector $\bfL$ mixes the polarization
modes of wavevectors $\bfl$ and $\bf{l^{\prime}}=\bfl -\bfL$. Taking the 
ensemble average of the CMB fields for a fixed $\alpha$ field, for $x\ne x'$ one gets
\begin{equation}
\langle x^\star(\bfl) x'(\bflp) \rangle_{\rm CMB} =  
                     f_{xx'}(\bfl,\bflp) \alpha(\bfL)\,,
\label{BasicDifference}
\end{equation}
here $x,x'\in\{T,E,B\}$; $f_{TB}=\tilde C_{l_1}^{T E}\cos 2\varphi_{\bfl_1\bfl_2}$, and 
$f_{E B}=2 [\tilde C_{l_1}^{E E}-\tilde C_{l_2}^{B B}]
\cos 2 \varphi_{\bfl_1\bfl_2}$. We have assumed that $\alpha<<1$ radian, which is an excellent approximation 
because current upper limits already set it on the sub-degree level~\cite{quad_parity,wmap7_cosmology}. The power spectrum is obtained by averaging over many realizations of the CMB, with $\alpha$ fixed. Note that the power spectrum is linearly proportional to $\alpha$ except for $x=x'=B$, where the $BB$ power spectrum is quadratic in $\alpha$.

If we also average over the rotation field then the above two point function vanishes for all $x,x'$ except for $x=x'$  for which  we are left with the rotation-induced CMB B-mode power spectrum,
\begin{eqnarray}
C^{BB}_L= 4\int \frac{d^2{\bf l'}}{(2\pi)^2} 
 C^{\alpha \alpha}_{l'} C^{EE}_{l''}\cos^2[2(\varphi_{\bfl''}-\varphi_{\bfL})]\,,
\end{eqnarray}
where $\bfL=\bfl' - \bfl''$. Note that no assumption has been made here as to the origin of this rotation, namely 
whether or not it is cosmological. In the literature, $\alpha$ is identified 
with the CB rotation angle (see \cite{wmap5_cosmology,quad_parity}). 

{\it Constant Rotation Case:} 
For the special case where CB is constant over the sky, i.e. the rotation angle is $l$-independent:
\begin{eqnarray}
a_{\ell m}^{E'} &=& a_{\ell m}^E \cos(2\alpha) - a_{\ell m}^B \sin(2\alpha) \nonumber \\
a_{\ell m}^{B'} &=& a_{\ell m}^E \sin(2\alpha) + a_{\ell m}^B \cos(2\alpha)\,,
\end{eqnarray}
and the power spectra become
\begin{eqnarray}
C_{\ell}^{'TB} &=& C_{\ell}^{TE} \sin(2\alpha) \nonumber \\
C_{\ell}^{'EB} &=& \frac{1}{2}\left( C_{\ell}^{EE} - 
C_{\ell}^{BB} \right) \sin(4\alpha) \nonumber \\
C_{\ell}^{'TE} &=& C_{\ell}^{TE} \cos(2\alpha) \nonumber \\
C_{\ell}^{'EE} &=& C_{\ell}^{EE} \cos^2 (2\alpha) + 
C_{\ell}^{BB} \sin^2 (2\alpha) \nonumber \\
C_{\ell}^{'BB} &=& C_{\ell}^{EE} \sin^2 (2\alpha) + 
C_{\ell}^{BB} \cos^2 (2\alpha)\,.
\end{eqnarray}

\section{Detectability of Cosmological Birefringence}
\label{sec:estimator}
 Following Ref.~\cite{2002ApJ...574..566H,kamionkowski_08,Yadav_etal_09,2009PhRvD..80b3510G,YSZ09}, an unbiased quadratic estimator $\hat{\alpha}_{xx'}(\bfL)$ for $\alpha(\bfL)$  
for  the CMB modes, $x x'=T B$ and $EB$ is 
\begin{eqnarray}
\hat \alpha_{xx'}({\bfL})& =&  N_{xx'}(L) \intl{1}
\, x(\bfl_1) x'(\bfl_2) 
\frac{f_{xx'}(\bfl_1,\bfl_2)}{C_{l_1}^{xx} C_{l_2}^{x'x'}}\,,
\label{eqn:estimator}
\end{eqnarray}
where $\bfL=\bfl_2  -\bfl_1$, and the normalization is given by
\begin{eqnarray}
N^{xx'}(L) = \Bigg[ \intl{1} f_{xx'}(\bfl_1,\bfl_2)
\frac{f_{xx'}(\bfl_1,\bfl_2)}{C_{l_1}^{xx} C_{l_2}^{x'x'}} \Bigg]^{-1} \,,
\label{eq:noise}
\end{eqnarray}
The fields $x(l)$ can be obtained from the observed map. Here, $C_{l_2}^{xx}$ and $C_{l_2}^{x'x'}$ are the observed power spectra         including the effects of both the signal and noise,
         \begin{eqnarray}
C_{l}^{xx}=C_{l}^{xx,theory}+\Delta^2_x e^{l^{2}\Theta^2_{\text{FWHM}}/(8\ln{2})}
\end{eqnarray}
where $\Delta_x$ is the detector noise and  $\Theta_{\text{FWHM}}$ is full-width at half-maximum (FWHM) of beamsize.
         
         The variance of the estimator can be calculated as
\begin{eqnarray}
\text{Var}(\tilde \alpha(\bfL))&=&\langle \tilde \alpha_{xx^\prime}(\bfL) \tilde \alpha^\star_{xx^{\prime}}(\bfL^{\prime})\rangle =
 N^2_{xx'}(L) \intl{1} \intl{3}
\, \langle x(\bfl_1) x'(\bfl_2)x(\bfl_3) x'(\bfl_4)\rangle \nonumber 
 \frac{f_{xx'}(\bfl_1,\bfl_2)}{C_{l_1}^{xx} C_{l_2}^{x'x'}}  \frac{f_{xx'}(\bfl_3,\bfl_4)}{C_{l_3}^{xx} C_{l_4}^{x'x'}}\,,
\nonumber\\
&=&(2\pi)^2 \delta(\bfL - \bfLp) \{ C^{\alpha \alpha}_L +N_{xx^\prime}(L)\}\,,
\label{eqn:variance}
\end{eqnarray}
where $\bfL=\bfl_2  -\bfl_1=\bfl_4  -\bfl_3$. In the last line, the first term is the desired CB power spectrum and second term is the estimator's noise for the reconstruction of CB. The expression for the noise is given by Eq.~(\ref{eq:noise}). The signal-to-noise ratio for detecting CB is given by~\cite{kamionkowski_08, Yadav_etal_09}.
\begin{eqnarray}
\left(\frac{S}{N}\right)^2&=& \sum^{l_{max}}_{2}\frac{f_{sky}}{2} 
    (2l+1) \left  ( \frac{C^{\alpha \alpha}_l}{N^{xx'}_l}\right )^2 \,,
\label{eq:s2n}
\end{eqnarray}
where $C^{\alpha \alpha}_L$ is the fiducial rotation angle power spectrum.  

\begin{figure*}[]
\centering
\includegraphics[scale=.80,angle=-0]{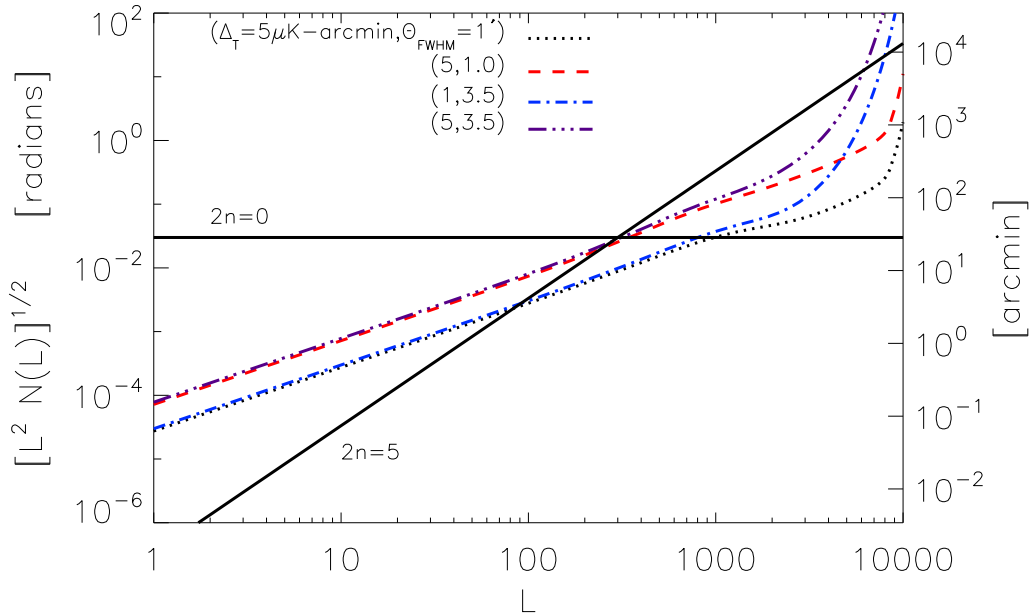}
\caption{Detectability of CB using the $EB$ estimator. 
We show the $EB$ estimator noise, $N(L)$, as given by Eq.~(\ref{eq:noise}), as a function of 
multipole $L$. We show the noise for four experimental setups. The noise, $\Delta_T$  in $\mu$K-arcmin and beam full-width at half-maximum $\Theta_{FWHM}$ -- units of arcmin, is labeled. For reference, the two solid black curves 
show the Faraday rotation  power spectrum from the stochastic magnetic field with ``causal spectrum'' ($2n=5.0$) and for a nearly scale-invariant spectrum ($2n=0.1$). 
}
\label{fig:FR_noise}
\end{figure*}

In Fig.~\ref{fig:FR_noise}, we show the quadratic $EB$ estimator noise as a function of multipole, $\ell$, for four experimental configurations. For reference we show two theoretical Faraday rotation power spectrum (black solid curves), one which corresponds to scale invariant magnetic field spectrum and other the for model with a causal stochastic magnetic field. Details on these models involving Faraday rotation can be found in Ref.~\cite{2011arXiv1106.1438P}.  
For the experiments considered here, the $EB$ estimator is the most sensitive, as will be demonstrated.

{\it A Constant Rotation Case:} For uniform 
rotation, $\bfL=\bfl_1-\bfl_2=0$ in Eq.~(\ref{eq:noise}), hence there 
is no mode mixing between different wavevectors. Although we do not show the ``monopole term'',  $L=0$, in Fig.~\ref{fig:FR_noise}, the estimator can also be used to find the detectability of uniform rotation. The signal-to-noise (S/N) for the detection of non-vanishing $C_{l}^{EB}$ is
\begin{eqnarray}
\big(S/N\big)_{EB}^{2}=\sum_{l}f_{sky}\frac{(2l+1)}{2}\frac{(C_{l}^{EB,obs})^{2}}{C_{l}^{E}C_{l}^{B}}
\end{eqnarray}
 The power spectrum in the numerator, $C_{l}^{EB,obs}$, is the observed spectrum, and the power spectra $C_{l}^{EE}$ and $C_{l}^{BB}$ in the denominator include the effects of noise and beam smearing (see Eq.~(10)). Similar expressions can be written for $(S/N)_{TB}$ by replacing $E$ by $T$. Note that for a cosmic-variance-limited experiment, $\big(S/N\big)_{EB}$ will exceed $\big(S/N\big)_{TB}$ because the 
cosmic variance of the temperature anisotropy is at least an order of magnitude larger than that of the E-mode polarization. 


\begin{figure}[]
\centering
\includegraphics[scale=.70,angle=-0]{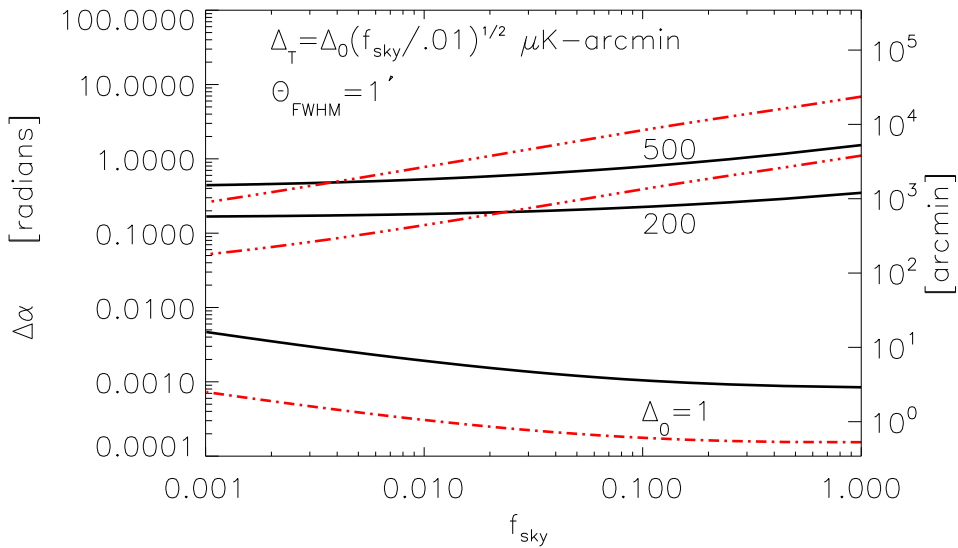}
\caption{Comparison of the $EB$ and $TB$ estimator for constant $\alpha$ case. 
We show forecasted uncertainties on constant $\alpha$ from the two estimators as a function of sky-fraction $f_{sky}$, for several choices of instrumental noise. The dashed-dotted red curves are for the $EB$ estimator while the solid black curves are for the $TB$ estimator. Clearly the $EB$ estimator outperforms the $TB$ estimator for the experiments with low enough noise such as EBEX~\cite{2010SPIE.7741E..37R}, CMBPol~\cite{CMBPol:inflation}, POLARBEAR~\cite{2011arXiv1110.2101K}, SPTPol~\cite{2009AIPC.1185..511M}, and SPIDER~\cite{2008SPIE.7010E..79C}. Note that for a given observing time, the noise $\Delta_T$ is proportional to $\sqrt{f_{sky}}$ and hence can be reduced by observing smaller sky fractions.
}
\label{fig:alpha_vs_fsky_error}
\end{figure}

Employing the conventional definition of the Fisher matrix, and assuming constraints from the $EB$ data, we obtain
\begin{eqnarray}
F^{EB}_{ij}=\sum_{l}f_{sky}\frac{(2l+1)}{2}\frac{\partial(C_{l}^{'EB})}{\partial\lambda_{i}}
\frac{\partial(C_{l}^{'EB})}{\partial\lambda_{j}}(C_{l}^{B}C_{l}^{E})^{-1}\,,
\label{eqn:fisher}
\end{eqnarray}
and the error on the parameter $\lambda_{i}$ is given by $\sigma_{i}=\sqrt{[(F^{EB})^{-1}]_{ii}}$, 
the $i'$th diagonal element of the square root of the inverse Fisher matrix. In the simplest case we consider, there 
is only a single parameter, the rotation angle $\alpha$, and the calculation is trivial.
A similar expression can be written for  the $TB$ estimator. It should be noted that assuming small instrumental noise,  such as EPIC~\cite{EPIC}, CMBPol~\cite{CMBPol:inflation, 2008arXiv0811.3918Z, CMBPol:lensing}, and COrE~\cite{core} the EB estimator outperforms the $TB$ estimator. This can be readily seen from the fact that $\sqrt{C_{l}^{T}C_{l}^{E}}\geq C_{l}^{TE}$, is the Cauchy-Schwarz inequality. In the other extreme, when the experiment is dominated by instrument noise , the $TB$ estimator performs the best. This is due to the fact that $C_{l}^{TE}>C_{l}^{EE}$ and the fact that 
$C_{l}^{E,det}=2C_{l}^{T,det}$. This result is also clear from  Fig.~\ref{fig:alpha_vs_fsky_error} where we compare the $EB$ and $TB$ estimator of various noise configurations.

\begin{figure*}[t]
\centering
\includegraphics[scale=.50,angle=-0]{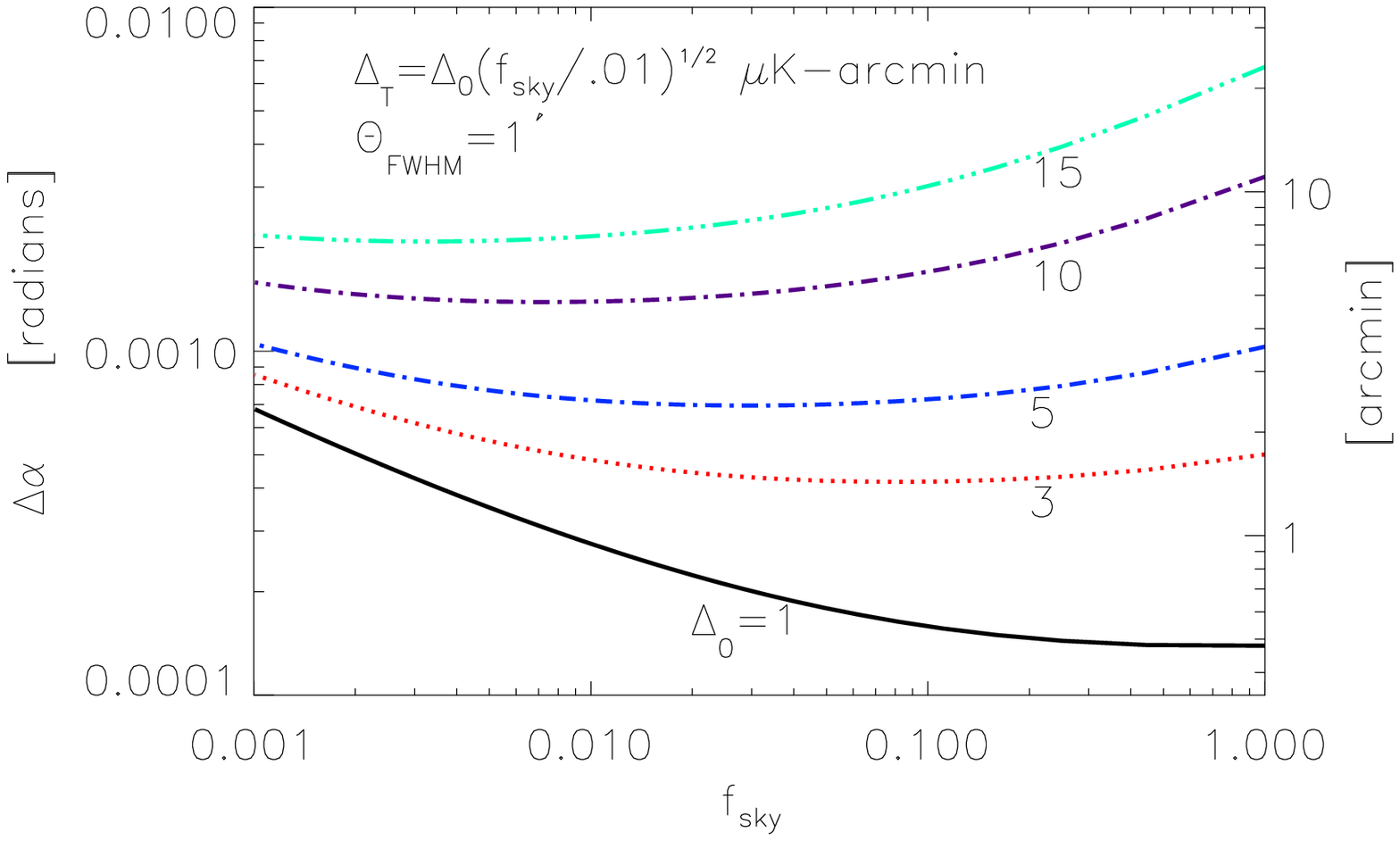}
\includegraphics[scale=.50,angle=-0]{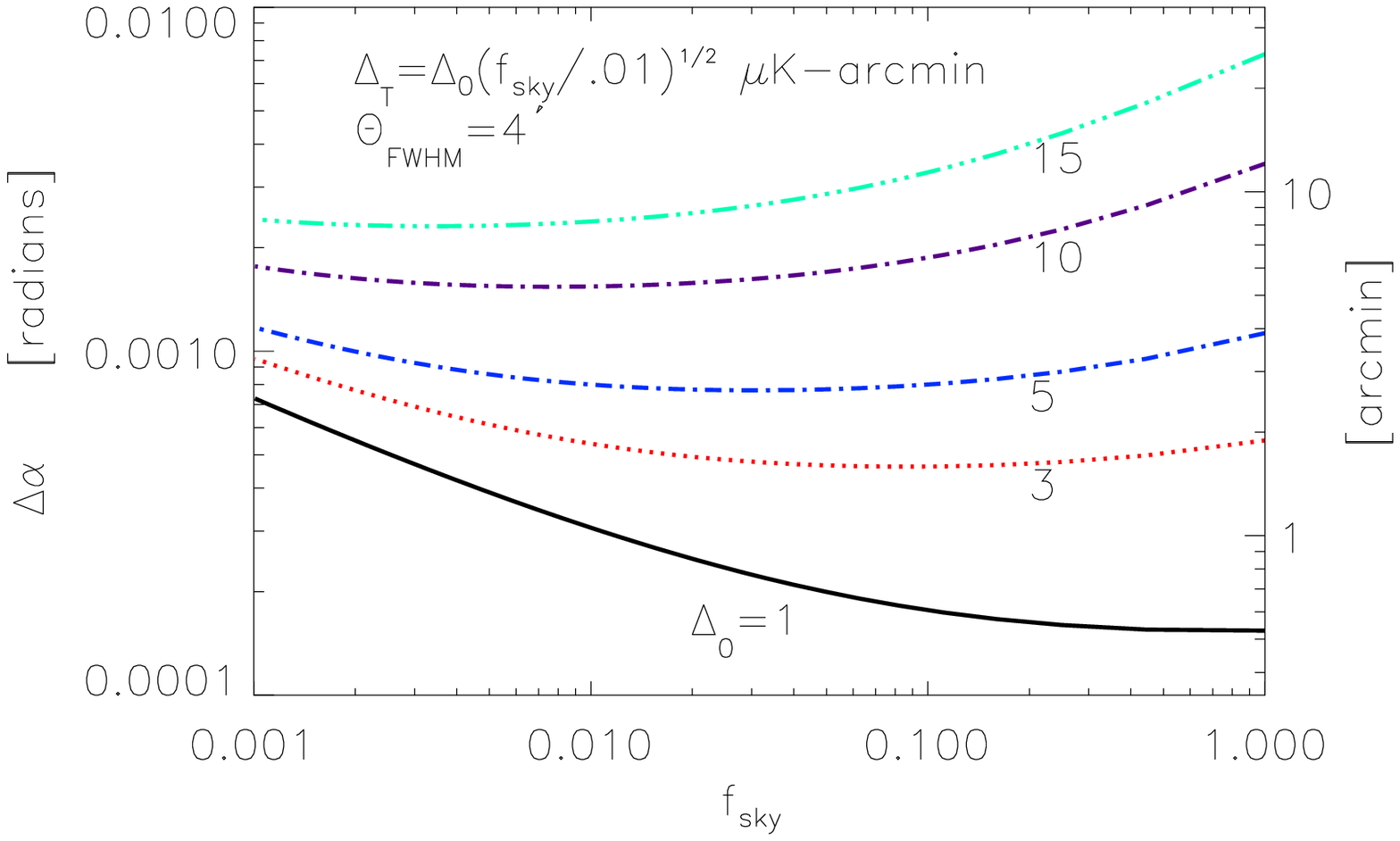}
\includegraphics[scale=.50,angle=-0]{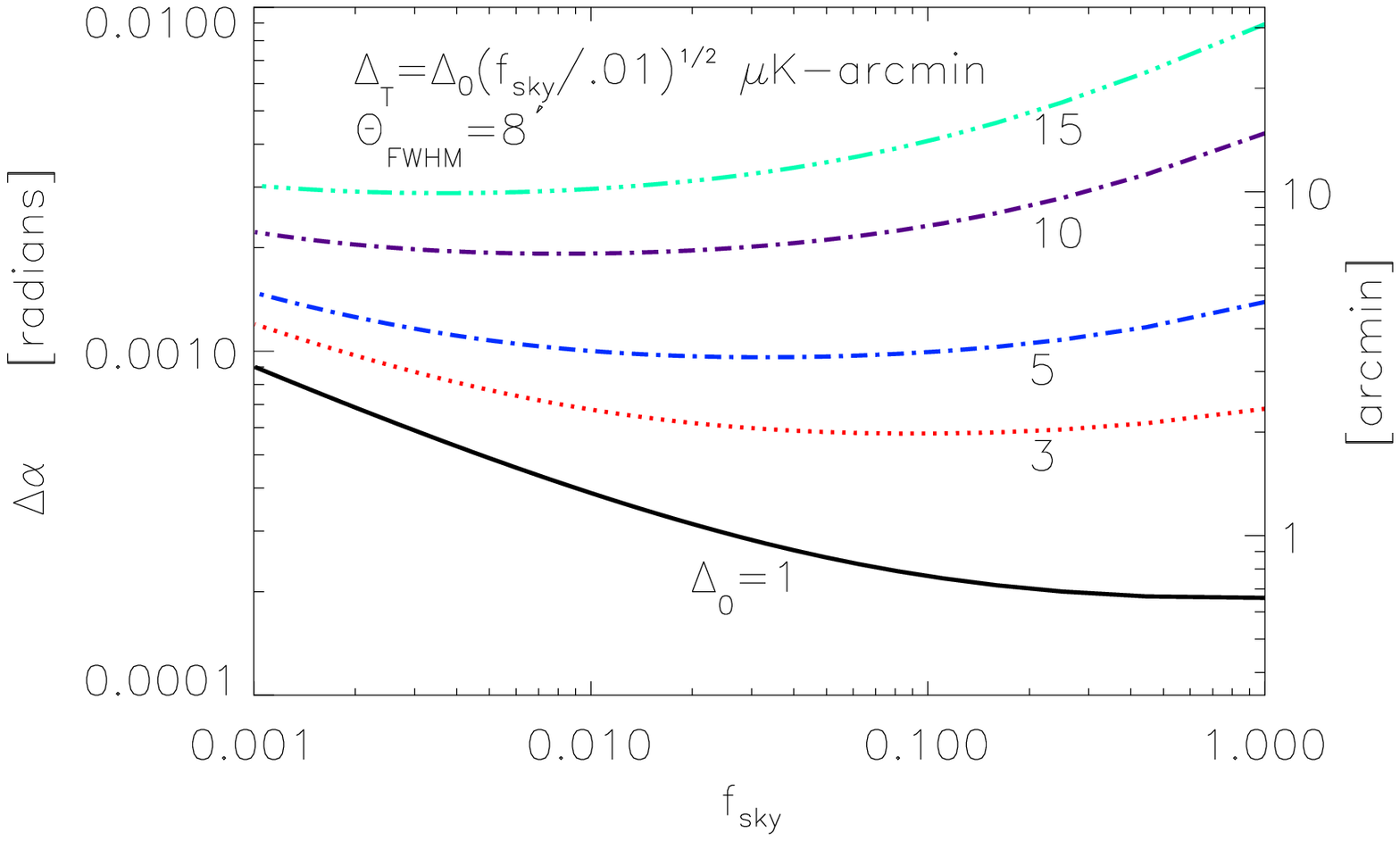}
\includegraphics[scale=.50,angle=-0]{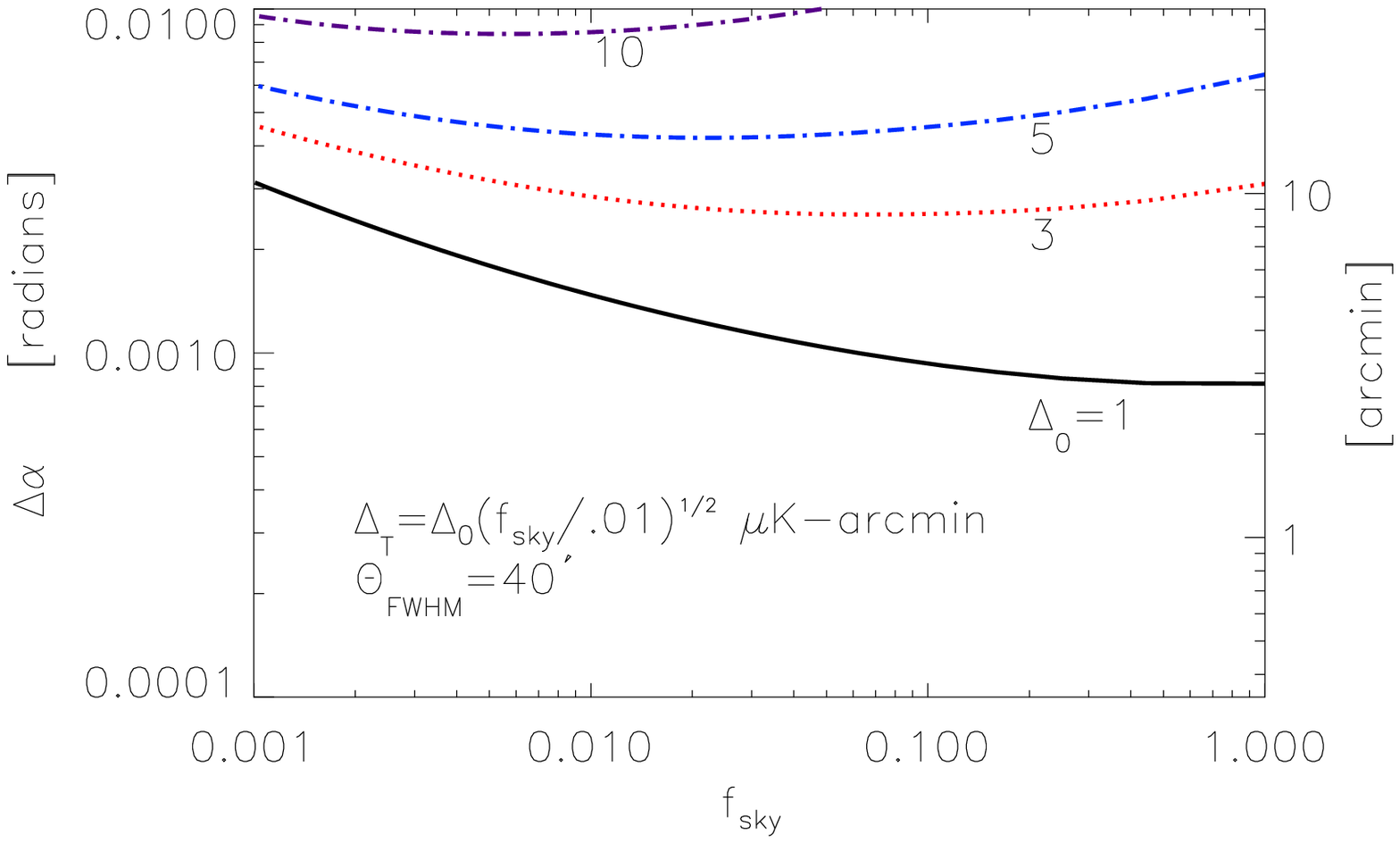}
\caption{ Forecasted uncertainties on constant $\alpha$ from the $EB$ estimator as a function of sky-fraction $f_{sky}$, for several choices of instrumental noise. The minima for the curves are a result of compromise between number of modes and noise $\Delta_p$. Note that the observing time was fixed.
}
\label{fig:alpha_vs_fsky}
\end{figure*}

Optimizing surveys for CB detection can be approached as follows: given the detector noise $\Delta_0$ for an experimental setup,  we can get $\Delta=\Delta_0(\frac{f_{sky}}{f^0_{sky}})^{1/2}$. Here $f^0_{sky}$ is a fiducial sky-fraction. Note that larger the $f_{sky}$, the shorter the observation time in a given sky direction, making the noise, $\Delta$, correspondingly higher.  In Fig.~\ref{fig:alpha_vs_fsky} we show how changing the sky-fraction, $f_{sky}$ changes the performance of the $EB$ estimator. In general there is a preferred $f_{sky}$ for which the error in CB is minimized. For the noise-dominated experiment it is preferred to have lower $f_{sky}$. For a nearly cosmic-variance-limited experiment it is always favorable to have larger $f_{sky}$. We discuss the physical reasons for this behavior in our results section.

\section{Requirement on Instrumental Systematics}
So far the CB detection prospects in the absence of systematics have been discussed. It has been shown that there are several instrumental systematics that can generate $EB$ and $TB$ correlations which also generate spurious $B$-mode polarization~\cite{Miller2009,YSZ09}. Although the CMB experiments considered in this work were designed to detect the very weak B-mode signal from inflationary gravitational waves~\cite{CMBPol:inflation,2011arXiv1110.2101K}, it has been shown~\cite{Miller2009} that this feature of these CMB experiments may be insufficient for an unbiased detection of CB. 

 Polarimeters such as  \emph{MaxiPol}~\cite{2007ApJ...665...42J}, \emph{Boomerang}~\cite{2003ApJS..148..527C}, \emph{BICEP}~\cite{2006SPIE.6275E..51Y}, \emph{QUAD}~\cite{2009ApJ...692.1221H} etc. difference intensity from bolometers sensitive to two orthogonal polarizations. Any differences between the two bolometers generates spurious $Q$ and $U$ signals. Furthermore, the spatial beams for each bolometer produces generally has some degree of ellipticity.
Following the formalism presented in Ref.~\cite{HHZ}, polarization systematics fall into two categories, one associated with the detector system which distorts the polarization state of the incoming polarized signal (hereafter ``Type I''), and another associated with systematics of the CMB signal due to the beam anisotropy (``Type II''). 
To first order, the effect of Type I systematics on the Stokes parameters can be written as~\cite{HHZ}
\begin{equation}
\delta [Q \pm i U](\bn) =
            [\calb \pm i 2 \rot](\bn)  [Q \pm i U](\bn) + [f_1 \pm i f_2](\bn)   [Q \mp i U](\bn)
            + [\gamma_1 \pm i \gamma_2](\bn) \cmb(\bn)\,,
\label{eq:pointsys}
\end{equation}
where $a(\bn)$ is a scalar field which describes the polarization miscalibration,
$\rot(\bn)$ is a scalar field that describes the rotation misalignment of the instrument,
$(f_1\pm if_2)(\bn)$ are spin $\pm 4$ fields that describe the coupling between two spin $\pm 2$
states (spin-flip), and $(\gamma_1\pm i \gamma_2)(\bn)$ are spin $\pm2$ fields that describe
monopole leakage, i.e. leakage from CMB temperature anisotropy to polarization. 

Similar to the Type I systematics, the effect of Type II systematics on
the Stokes parameters can be written as~\cite{HHZ}
\begin{equation}
\label{eq:localmodel}
\delta[Q \pm i U](\bn;\sigma) = \sigma {\bf p}(\bn) \cdot \nabla [Q \pm i U](\bn;\sigma)
+ \sigma [d_1 \pm i d_2](\bn) [\partial_1 \pm i\partial_2] \cmb(\bn;\sigma)
 + \sigma^2 q(\bn) [\partial_1 \pm i \partial_2]^2 \cmb(\bn;\sigma)\, ,
\end{equation}
where the systematic fields are smoothed over the coherence size $\sigma$. The spin $\pm 1$ fields, $(p_1\pm ip_2)$ and $(d_1\pm id_2)$, describe pointing errors
and dipole leakage from temperature to polarization, respectively, 
and $q$ is a scalar field that represents quadrupole leakage~\cite{HHZ}, e.g. beam ellipticity.

We  now discuss how systematics contaminate the rotation estimator.  We start by defining a matrix $F^{{\alpha}{\cal D}'}_\ell$ for each multipole $\ell$,
\begin{eqnarray}
F^{{\alpha}{\cal D}'}_\ell &=& \intl{1}f^{{\alpha}}_{EB}(\ell_1,\ell_2)
({\bf C}^{-1})^{EE}_{\ell_1}f^{{\cal D}}_{EB}(\ell_1,\ell_2)
({\bf C}^{-1})^{BB}_{\ell_2} \,, 
\end{eqnarray}
where  ${\bf \ell}=\bf {\ell}_1 -{\bf \ell}_2$, and ${\cal D}$ run over  all the 11 systematics and lensing, $\{a,\rot, \gamma_1, \gamma_2, f_1, f_2, d_1, d_2, q, p_1, p_2, \phi\}$. The filters $f^{\cal D}(\ell_1,\ell_2)$ are given in Ref.~\cite{HHZ,YSZ09}. Our estimator, in the absence of any distorting systematics field, is unbiased, meaning that
\begin{eqnarray}
\big\langle\hat {\alpha}(\bfL)\big\rangle_{CMB}={\alpha}(\bfL)\,.
\label{eqn:unbiased}
\end{eqnarray} 
However, in the presence of multiple systematic effects the estimator as constructed in Eq.~(\ref{eqn:estimator}) may not be unbiased. In general Eqn.~(\ref{eqn:unbiased}) is modified as:
\begin{equation} 
\langle \hat {\alpha}(L)\rangle_{CMB} = {\alpha}(L) + \frac{ \sum_{{\cal D}} F_L^{{\alpha}{\cal D}} {\cal D}'(L)}{F^{{\alpha}{\alpha}}_L} \,,
\label{fisher_bias}
\end{equation}
where $\hat {\alpha}$ is an estimate of ${\alpha}$ in the presence of multiple systematics. The level of bias depends on the Fisher matrix, $F_L^{{\alpha}{\cal D}}$, and the amplitude of systematics, ${\cal D}$. 

Systematics induced $B$-mode power spectrum is given by 
\begin{equation}
\langle B(\bfl_1) B(\bfl_2) \rangle_{\rm CMB, \cal{D}} = (2\pi)^2 \delta(\bfl_1+\bfl_2)
\begin{array}{cc}
\displaystyle \int  \frac{d^2\bfl'}{(2\pi)^2} C^{{\cal D}{\cal D}}_{l'} \tilde C^{xx}_{l''} W^{B}_{\cal D}(\bfl',\bfl'') W^{B}_{\cal D}(\bfl',\bfl'') \,,
\end{array}
\label{eqn:cbb}
\end{equation}
where $\bfl''=\bfl_1-\bfl'$; $x=E$ for systematics with E to B leakage, (${\cal D}={a,\rot,(f_1,f_2),(p_1,p_2)}$);  and  $x=T$ for systematics with T to B leakage (${\cal D}={q,\rot,(\gamma_1,\gamma_2),(d_1,d_2)}$) . We have assumed zero primordial B-modes for our fiducial model. The window function, $W^{B}_{\cal D}(\bfl_1,\bfl_2)$, for each of the systematics, $\cal D$, are given in Ref.~\cite{HHZ,YSZ09}.

In Table~\ref{tab:systematics} we compare the requirement on instrumental systematics for the measurement of CB either via the B-modes or by using the $EB$ estimator. In the table, the first column lists the systematics considered. In column 2-5 we show the requirement from rotation induced $B$-mode power spectrum. This requirement was derived by demanding that the systematics induced $B$-mode power spectrum (as given by Eq.~(\ref{eqn:cbb}) is 10 times smaller than the rotation induced $B$-mode power spectrum detectable by that experimental setup. In column 6-9 give the minimum $rms$ level of each systematics ${\cal D}$ such that the bias generated by that systematic for CB estimation is 10 times smaller than the rotation field. To be more precise, we require that the underlying CB power spectrum $C^{\alpha \alpha}_\ell$ is 10 times smaller than the bias contribution to power-spectrum arising from second term on the right hand side of Eq.~(\ref{fisher_bias}).  We used $L=100$ for this requirement however the requirement for $L=0$ (the monopole, or constant rotation case) is comparable. We show the numbers for several experimental configurations. Note that the systematics  fields $a,\gamma_1,f_1,f_2,d_2,q$ and $p_2$ are uncorrelated with rotation $\alpha$ and hence do not bias the estimator.  For the remaining systematic fields $\gamma_2,d_1$ and $p_1$ the requirement is not particularly stringent when compared to the requirement for extracting rotation from the $B$-mode power spectrum. The $rms$ requirement on the systematics depends on the experimental  configuration and for experiments in consideration in Table I. Even the most stringent requirement is around $5\%$, which is easily achievable with current experiment designs. We note that for the simple frequency-independent model shown earlier, constant CB is completely degenerate with the differential rotation beam systematic $\rot$~\cite{Miller2009}. Therefore measuring anisotropic $\alpha(\bn)$~\cite{2012arXiv1206.5546G} is \emph{less sensitive} to systematics contamination.

Throughout our analysis we have assumed systematics to be isotropic Gaussian fields with power spectra of the form
\begin{equation}
C_l^{{\cal D}{\cal D}} = A^2_{\cal D} \exp(-l(l+1)\sigma_{{\cal D}}^2/8\ln2), \label{eqn:css}
\end{equation}
i.e. white noise above a certain coherence scale $\sigma_{{\cal D}}$. The parameter $A_{\cal D}$ characterizes the $\it rms$ of the systematics field ${\cal D}$. For our numerical calculations we have assumed $\sigma_{\cal D}=120'.$ 
 
   \begin{table} [t]
\begin{tabular}{c||c|c|c|c||c|c|c|c}
 &  \multicolumn{8}{c} {Requirement}  \\ 
\cline{2-9}
 &  \multicolumn{4}{c||} {B-mode Spectrum} & \multicolumn{4}{c} {EB Estimator}  \\ 
\cline{2-9}  Systematics &  E1 ($\Delta_T=1,1'$)& E2 ($5,1'$) & E3 $(1,3.5')$& E4 $(5,3.5')$&  E1 $(1,1')$ & E2 ($5,1'$)& E3 $(1,3.5')$& E4 $(5,3.5')$\\
\hline $a$ & $1.1\times 10^{-3}$ & $2.8\times 10^{-3}$ & $1.2\times 10^{-3}$  & $3.1\times 10^{-3}$ & -  & - & - &  -\\
\hline $\gamma_1$ & $1.1\times 10^{-5}$ & $2.7\times 10^{-5}$  & $1.1\times 10^{-5}$ & $3.0\times 10^{-5}$ & - & - & - & -\\
 \hline $\gamma_2$ & $8.6\times 10^{-6}$ & $2.2\times 10^{-5}$ &$9.4\times 10^{-6}$ & $2.5\times 10^{-5}$ & 0.10  & 0.08 & 0.1 & 0.08  \\
 \hline $f_1$ & $1.1\times 10^{-3}$ & $2.9\times 10^{-3}$ & $1.2\times 10^{-3}$ & $3.2\times 10^{-3}$ & - & - & -&  -\\
 \hline $f_2$ & $1.0\times 10^{-3}$ & $2.6\times 10^{-3}$ & $1.1\times 1o^{-3}$ & $2.9\times 10^{-3}$ & -  & - &- &-  \\
\hline  $d_1$&  $1.2\times 10^{-4}$ & $3.2\times 10^{-4}$ & $7.1\times 10^{-5}$ & $1.8\times 10^{-4}$ & 0.1  &0.56 &0.04 & 0.24  \\
 \hline $d_2$ &  $1.3\times 10^{-4}$ & $3.5\times 10^{-4}$ & $7.8\times 10^{-5}$  & $2.0\times 10^{-4}$ & -& - & -&-\\
\hline  $q$ &  $1.4\times 10^{-3}$ & $3.6\times 10^{-3}$ & $4.3\times 10^{-4}$ & $1.1\times 10^{-3}$ & - & - & - &  -\\
 \hline $p_1$& $2.0\times 10^{-2}$ & $5.3\times 10^{-2}$ & $1.2\times 10^{-2}$ & $3.06\times 10^{-2}$ &0.16  & 0.4& 0.05 &  0.13\\
 \hline $p_2$ & $1.2\times 10^{-2}$ & $3.3\times 10^{-2}$ & $7.3\times 10^{-3}$ & $1.9\times 10^{-2}$ & - & - &- & - \\

\hline
\end{tabular}
\caption{Requirement on instrumental systematics for the measurement of cosmic birefringence. The first column lists the systematics considered while remaining columns show the minimum-systematics {\it rms}  requirement for different experimental setups (E1-E4). Experimental noise $\Delta_T$ in $\mu$K-arcmin and beam full-width at half-maximum, $\Theta_{\text FWHM}$, for each experiment is provided in the round brackets.  In column 2-5 we show the requirement from rotation induced $B$-mode power spectrum. This requirement was derived by demanding that the systematics induced $B$-mode power spectrum is $10$ times smaller than the rotation induced B-mode power spectrum detectable by that experimental setup. Columns 6-9 shows the requirement for the $EB$ quadratic estimator. This requirement was derived using Eq.(\ref{fisher_bias}), where we demand that the spurious power spectrum contamination to the rotation $EB$ estimator (coming from the second term at the right hand side of Eq.(\ref{fisher_bias})) is 10 times smaller than the fiducial underlying underlying rotation power spectrum.}
\label{tab:systematics}
\end{table}

{\it Beam Systematics for a Constant CB Case:}
We discussed the requirements on systematics, however we can marginalize 
over unknown beam parameters. This is possible due to the fact that e.g. beam ellipticity and 
beam offset have different $L$-dependence than CB and can therefore be distinguished from a pure 
CB. In other words, the Fisher matrix is regular and invertible. However, pixel rotation is completely degenerate with $\alpha$ and we therefore can offer no method of disentangling it 
from the measured $\alpha$ other than a precise calibration with polarized sources~\cite{KSZ12}. It turns out that 
including the beam ellipticity and beam offset effects in our Fisher matrix estimation is surprisingly easy. The analytic model for beam systematics considered by Ref~\cite{Shimon2008}
gives the systematic $TB$ and $EB$ spectra from the combined effect of beam offset $\rho$ and ellipticity $e$
\begin{eqnarray}
C_{\ell}^{TB,sys}&=&A_{1}l^{2}C_{\ell}^{TT}\nonumber\\
C_{\ell}^{EB,sys}&=&A_{2}l^{4}C_{\ell}^{TT}
\end{eqnarray}
where the two new free parameters $A_{1}$ and $A_{2}$ are certain combinations of $e$ and $\rho$.
The Fisher matrices in Eq.~(\ref{eqn:fisher})  now become two dimensional. In Table~\ref{tab:alpha} we show $1\sigma$ error on constant CB, $\alpha$ from the $EB$ and the $TB$ estimator with and without the systematics parameters marginalized. It is clear from Table I that for various experimental configurations, the systematics increases the error by only $10-20\%$.

\section{Results and Discussion}
In Fig.~\ref{fig:FR_noise} we show the the minimum detectable spatially dependent CB. We show the $EB$ quadratic estimator noise as a function of multipole $L$ for several experimental configurations. For a given multipole $L$, the noise level sets the minimum detectable signal for that multipole.  
For reference we have also shown two fiducial theoretical Faraday rotation power-spectra, for a scale-invariant and causal stochastic magnetic field~\cite{2011arXiv1106.1438P}. It is clear from the plot that for scale invariant case largest contribution to the  $(S/N)$ comes from large-scale (low $L$), while for causal magnetic fields, small scales (large $L$) contribute the most.

In Fig.~\ref{fig:alpha_vs_fsky_error} we compare the $TB$ and $EB$ estimator for the constant CB case. It is clearly seen that the $EB$ estimator results in better 
constraints on $\alpha$ for current and future polarization experiments. We also show curves for relatively large experimental noise cases $\Delta_T=200, 500 \mu$K-arcmin to show that, for experiments with larger noise levels, the $TB$ estimator  performs better.

In Fig.~3 we study the optimization of polarization experiments seeking to constrain CB. We show the minimum detectable rotation angle as a function of the sky fraction, $f_{sky}$, for several choices of instrument detector sensitivity. Here the observing time was  fixed. For smaller $f_{sky}$ the noise will be lower but at the cost of smaller number of modes. While if $f_{sky}$ is large, we will have more modes at the cost of larger noise.  This compromise between the number of modes and noise is visible in the plot in the form of preferred minima in the curve $\Delta \alpha$-versus-$f_{sky}$. 

In Table~\ref{tab:systematics} we study the requirements on instrumental systematics to be able to utilize a given experiment for detecting or constraining CB. Here we have considered 11 systematics. For each systematic we ask, how small their $rms$ fluctuation must be such that they do not affect detection of CB using (1) the $B$-mode power spectrum (2) $EB$ estimator. The conclusion of this analysis tells that $EB$ estimator is much less prone to instrumental systematics, i.e. the requirement on $EB$ systematics effects are much weaker than for the B-mode spectra. In fact there are several systematics ($a,\gamma_1,f_1,f_2,d_2,q$ and $p_2$) which do not affect the estimation of rotation with $EB$ estimator while all the systematics affect the estimation using the $B$-mode spectrum. Even for the systematics which do affect $EB$ estimator, the requirements on systematics $rms$ control is three-orders of magnitude smaller than the corresponding requirement from $B$-modes.  We also note that lensing deflection is same as the pointing systematics parameter  $p_2$. We show in Table I that $p_2$ is orthogonal to CB and does not bias the CB estimator. Hence lensing also does not bias the CB estimator. 

In Table II we give our forecasted Fisher uncertainties on constant CB $\alpha$. The experimental configurations used to get the constraints are also shown in the first three columns. 
 It is seen from the table II that the $EB$ estimator gives constraints $\sim 3$ times better than those obtained from employing 
the $TB$ estimator this is due to the fact that the experiments considered here are near cosmic-variance-limited and 
that $C_{l}^{TT}$ is $\sim 10$ times larger than $C_{l}^{EE}$ on the range of multipoles considered 
($l_{max}=3000$ in our calculation). It is also apparent from Table II that adding in the beam systematics does not 
degrade our constraints by more than 20\% due to the fact the beam-systematics-induced $TB$ and $EB$ are 
only weakly degenerate with pure CB (different $L$-dependence and the fact that the experiments we 
consider cover a sufficiently wide multipole range to allow separation of systematic $TB$ and $EB$ from cosmological 
birefringence). One could add additional free parameters but our results are 
not expected to significantly change because the $L$-dependence is different enough to allow separation between CB and 
beam systematic effects.

\begin{table} 
\begin{tabular}{c |c|c||c|c|c|c}
\hline $\Delta_T$ [$\mu$K-arcmin] & $f_{sky}$ & $\Theta_{\text FWHM}$ [arcmin]& $\Delta\alpha_{TB}$ [rad]& $\Delta\alpha_{TB+sys}$ [rad]& $\Delta\alpha_{EB}$ [rad]& $\Delta\alpha_{EB+sys}$ [rad]\\
\hline 1.0 & 0.1 & 1.0 & 0.27E-3 &0.30E-3& 0.87E-4 & 0.10E-3 \\ 
\hline 5.0 & 0.1 & 1.0 & 0.66E-3 &0.69E-3& 0.22E-3 & 0.25E-3 \\ 
\hline 5.0 & 0.1 & 3.5 & 0.71E-3 &0.74E-3& 0.24E-3 & 0.28E-3 \\ 
\hline 1.0 & 0.1 & 3.5 & 0.29E-3 &0.32E-3& 0.94E-4 & 0.11E-3 \\  %
\hline 1.0 & 0.01 & 1.0 & 0.88E-3 &0.96E-3 & 0.27E-3 & 0.32E-3 \\ 
\hline 5.0 & 0.01 & 1.0 & 0.21E-2 &0.22E-2& 0.72E-3 & 0.82E-3 \\ 
\hline 1.0 & 0.01 & 3.5 & 0.95E-3 &0.10E-2& 0.30E-3 & 0.35E-3 \\ 
\hline 5.0 & 0.01 & 3.5 & 0.23E-2 &0.24E-2& 0.79E-3 & 0.91E-3 \\ 

\hline
\end{tabular}
\caption{Projected constraints from various experimental configurations on the constant cosmic rotation $\alpha$ .  Experimental specifications are shown in column 1-3. Column 4-7 show  the expected uncertainties on $\alpha$ 
from both the $TB$ and $EB$ estimators, without and with systematics marginalized. It is clear that the $EB$ estimator yields tighter constraints on $\alpha$. }
\label{tab:alpha}
\end{table}

\section{Conclusions}
Any mechanism capable of converting 
$E$- to $B$-mode polarization will necessarily leak the  $TE$ and $EE$ correlations 
to $TB$ and $EB$, respectively. 
Non-vanishing $TB$ and $EB$ could be used, for example, to monitor residual 
systematics during data processing~\cite{YSZ09, Miller2009}, to constrain Faraday rotation, or to constrain parity violating physics. In this paper we presented a simple analytic approach to the optimization of ground-based CMB polarization observations capable of CB detection. We studied and compared the detection capabilities using the $EB$ correlations, $TB$ correlations, and the B-mode power spectrum. We find the $EB$ correlation based estimator to be the best, both in terms of having least noise and also suffering least from systematic effects. 
Special care should be given to systematics because they can generate spurious $EB$ correlations in the CMB - widely considered as smoking gun for CB. We demonstrated that this is probably not a significant issue when one performs a 
global parameter estimation, leaving beam parameters as nuisance parameters in the analysis.
We find that that in general there is a preferred sky fraction $f_{sky}$ for which the errors on CB are minimized. 
However for experiment with noise $<5 \mu$K-arcmin, 
very deep observing mode will result in weaker constraints on CB and that increasing the observed 
sky area typically exhausts the capacity to constrain CB. However, adopting such a wide sky coverage will come at the expense of the very high  sensitivity required for the inflation-induced B-mode -- the tradeoff between these two scientific goals is clear and in order to optimize future experiments it is desirable to define a metric for the successful measurement of both signals. Complications from astrophysical foregrounds were not considered in this work. It is our hope that future CMB observations will provide crucial information of the emission mechanisms that could produce spurious $TB$ and $EB$ correlations in the CMB. In any case, these foregrounds have very different spectral radiance behaviors than the CMB and so by using multi-frequency observation one may hope to remove their contribution, at least in part. 
In addition, their spatial profile and possibly their clustering properties could be used to further identify these non-cosmological contributions and separate them out from the data.

%

\section*{Acknowledgements}
{A.P.S.Y. acknowledges support from  NASA grant number NNX08AG40G. MS was supported by the James B. Ax Family Foundation.}

\end{document}